# Sitting on a Gold Mine: The Story of the Process Industry's Automatic Formation of a Digital Twin

ENHANCE workshop: Enabling Technologies and Dependability in Cyber-Physical Systems

**Mohammad Azangoo, Seppo Sierla, Valeriy Vyatkin**
Department of Electrical Engineering and Automation
Aalto University
Espoo, Finland
{mohammad.azangoo, seppo.sierla, valeriy.vyatkin}@aalto.fi

September 20, 2023

## Abstract

The use of a software tool chain to generate Digital Twins (DTs) automatically can speed up digitization and lower development costs. Engineering documents and system data are just two examples of source information that can be used to generate a DT. After proposing a general plan for semi-automatic generation of a DT for a process system, this work describe our efforts to extract necessary information for the generation of a DT of a process system from existing information in a factory floor like piping and instrumentation diagrams (P&IDs). To extract initial raw model data, techniques such as image, pattern, and text recognition can be used, and then an intermediate graph model can be generated and modified based on requirements. In order to increase the system's adaptability and reliability, this research will delve deeper into the steps involved in creating and manipulating an intermediate graph model.

## 1 Introduction

During the industry 4.0 era, demands for advanced software and machine interaction increased, and new domains like Cyber-Physical Systems (CPS), Digital Twin (DT), and Industrial Internet of Things (IIoT) became more popular. The main concept behind all of these topics is integration between the physical world and virtual systems in a way that improves quality, efficiency, and safety. In CPS, physical and software components are intensively interconnected. The software side of CPS can support hardware for better performance and more reliability. Similarly, a DT has a digital or virtual representation of a physical system that should be in contact with the physical system [Tao et al., 2019]. As a main software infrastructure, a DT should have a simulation model of a system, while the software side of a CPS can be designed and developed freely in any applicable format to achieve different tasks like controlling, monitoring, safety, and decision-making. So, generally, DTs can be considered as a subset of CPS, which can lead to realizing CPS.

The transition from a traditional form of production to the industry 4.0 compatible format could be very challenging. For example, using a DT needs an updated model of the plant; generation of the model needs cost, time, and human experts. However, for industry owners whose mills have a large number of engineering documents and data repositories, it's not a big deal. They are definitely sitting on a gold mine. They own something very valuable without realizing it. Engineering documents and production data history (gold mines) are rich in information (gold) that can be extracted to create DT simulation models (24-karat pure gold). Automatic tools for different steps like searching, digging, extracting, transferring, processing, and creating final products using pure gold would make gold production more affordable and efficient. Similarly, using an automatic software chain to generate a DT can speed up the process of digital transformation.

We are working on the subject of the automatic generation of DTs to discover potential opportunities for affordable transformation from traditional to digitalized production. So far, our focus has been on the domain of the process industry. The main idea for the automatic generation of DTs in the process industry, from our point of view, has been published as a general road map for brownfield process systems to the semi-automatic generation of DTs [Sierla et al., 2021]. It would be perfect if we could make all steps toward the automatic generation of DTs automatic, but in reality (at the moment), it is always needed to receive a level of support from human experts. The generation of DTs is a multidisciplinary task which needs expertise from different fields like computer science, computer vision, modeling and simulation, systems and control, and also process engineering (the domain that the plant and DT belong to). The final product, for the automatic DT generation, will be modular and include a set of diverse but compatible software modules that can work together to achieve a certain goal for the generation of a specific DT. For example, Sierla et al. [2020a] presented a chain of software tools to generate a steady state DT for a laboratory scale water process plant in a semi-automatic way. Their tools were about the extraction of a graph model, model modification, and consideration of simulation software related rules. This paper presents a framework for the automatic generation of DT in the process system and addresses steps to make the solution more comprehensive and more automated. In addition, we will go into further detail regarding the intermediate graph model, including the type of flexibility and opportunities that it will allow via the use of Graph Theory techniques.

## 2 Automatic generation of a digital twin

In the process systems, 3D CAD models and technical documentation such as P&IDs include the necessary information for automatically generating a simulation model [Sierla et al., 2020b]. It is needed to develop algorithms for recognizing main items in the system, such as mixers, heat exchangers, and pumps, from P&IDs and 3D CAD models, in order to automatically combine this information into a unified digital model. Different sources of information can be compared by transferring them to a similar level of abstraction and creating a matched and unified model [Sierla et al., 2020c].

Using software solutions can guarantee the affordable formation of DTs. Figure 1 depicts the fundamental configuration of the system for the generation of a DT by linking various software tools and technologies. This structure can extract relevant data from available sources of information like P&IDs, history of the process, and 3D information and build a machine-readable system model in an operational process system. Then, an intermediate graph-based model with a lot of information about different process components, flows, and component attributes can be made by combining the models from different sources of information. This makes the intermediate model comprehensive and adaptable, and it can be used for a wide range of applications. Next, the intermediate model needs to be improved so that the final simulation model can be made and the simulation software can get any extra information it needs. Finally, a DT can be made by connecting a simulation model in a simulation software to the process system.

### 2.1 Graph model

We can use graph theory to deal with the complexity of technical modeling. The graph model can help us convert complex models into a collection of elements and connections in addition to data attributes. This will give us a lot of flexibility and benefits. A graph is fundamentally very simple in comparison to many other mathematical models. There are also many tried and proven algorithms for optimization, consistency checks, and other tasks that can be applied to graph models. Graphs can be used to look up relationships or logic between elements, as well as to check if the model meets specifications. In addition, graphs allow us to create precise, imaginative relationships between elements, and it would be easy to visualize and show the system's topology, which would be very useful for DT use cases. Another important factor is that most studies in the same field use graph models on a regular basis [Son et al., 2015, Sierla et al., 2020b].

The intermediate graph model is at the core of the presented configuration for generation of a DT. Figure 2 shows an Unified Modeling Language (UML) Class Diagram for a graph model of a process system created for a simulation software. This model contains all of the nodes, connections, and features needed to build a simulation model and DT required for that simulation software. Attributes required in the graph model can vary depending on the level of fidelity and form of simulation model, simulation software requirements, and available input data. Steady state simulation, which needs a lower fidelity simulation model, ignores time-dependent behavior and is mostly utilized when planning a plant or retrofit to determine yields, raw material consumption, and emissions [Sierla et al., 2020a]. Dynamic simulation, which needs a higher fidelity simulation model, is used to assess how a process will respond to transitory occurrences such as a change in set-point, a valve closure, or an equipment malfunction [Azangoo et al., 2021]. As a result, developing a dynamic simulation model necessitates a large amount of extra source data that is not required for steady state models. Everything connected to the control system is an example of such information. Also, different types of input files can provide different levels of information. For example, P&ID files can provide information about the general connection of equipment as well as control signals, whereas 3D CAD files can provide exact geometry and topology of the system but not control signals In addition, simulation software and the environments in which a digital twin is operated may differ, necessitating differing requirements to be matched with the graph model. In this case, an interface is required to convert the graph model into a format that can be used by simulation software. It may be necessary to add more nodes or edges to a model to make it compliant with simulation software's requirements, as seen in Figure 2.

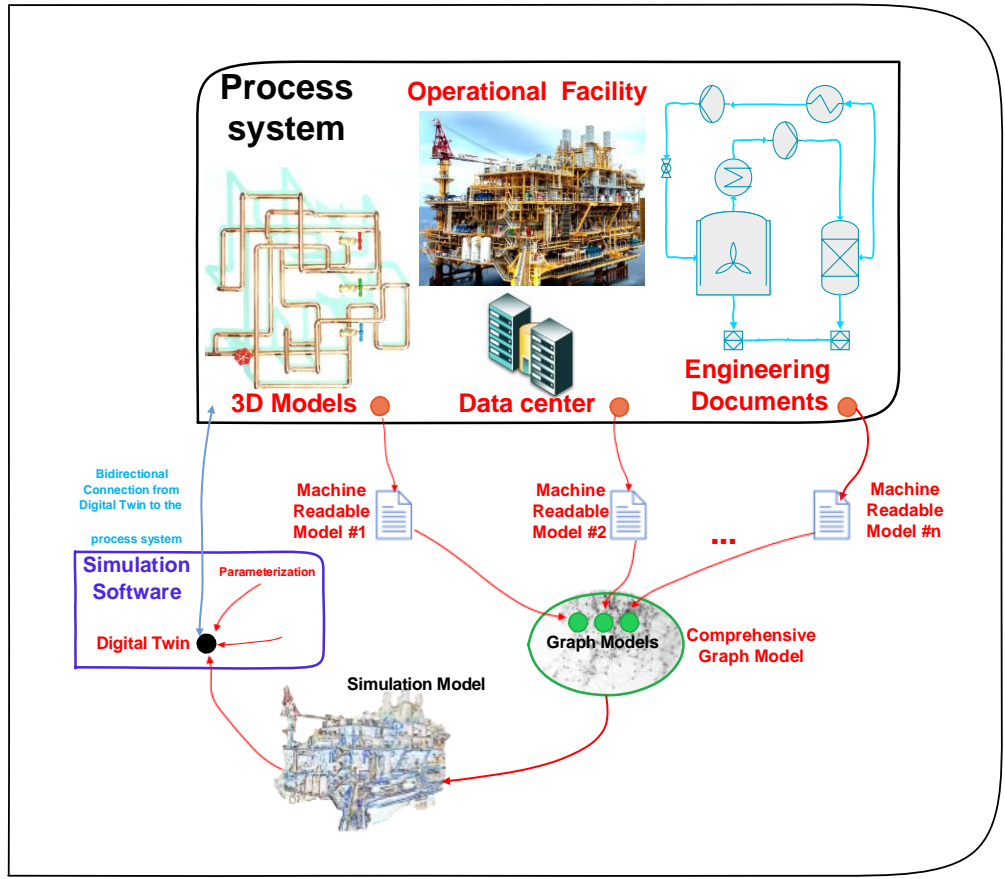

Figure 1: Structure and connection between subsystems for generation of a DT for an operational process system.

### 2.2 Attributes for a graph model

Graph models consist of nodes, which represent actual system elements such as tanks and pumps, connected by edges, which represent pipelines that connect the elements together in a process system. The graph's nodes and edges can be filtered to get the desired level of fidelity. For example, since the steady state simulation does not require sensor data, the related nodes and edges can be removed. If we understand the nature and characteristics of the nodes and edges, we may adapt the graph to meet the requirements. Hence, as seen in Figure 2, attributes must be utilized to represent all nodes and edges. Some of the most important attributes that should be considered for nodes in their data structure are type, tag, position in the original document or geometry information, number of input or output flows (nozzles can be considered separately as nodes), and specific characteristics related to different types of nodes, like volume of a tank or max flow of a pump. Edges in a graph can additionally include attributes like source and target nodes, flow materials, and size information.

## 3 Conclusion and future work

In this paper, we set out a general plan for getting the information needed to make a digital twin of a process system from existing data, like engineering documents. Techniques like picture, pattern, and word recognition can be used to get the raw data for the model. Then, an intermediate graph model can be built and modified based on the needs. After

SCEES

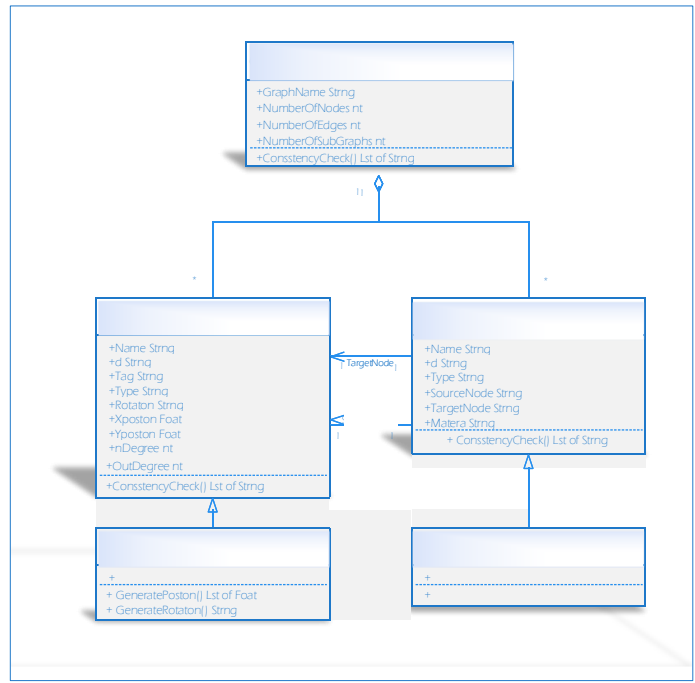

Figure 2: An UML class diagram for representation of the graph model adapted from [Sierla et al., 2020a].

proposing the plan to make a DT for a process system semi-automatically, this work moved on to show how to make an intermediate graph model to develop the plan more flexible and consistent. The intermediate graph model should be modified and used in future work for the generation of DTs or other applications based on end-user needs.